\documentclass[twocolumn,english,aps,prl,showpacs,superscriptaddress]{revtex4-1}
\usepackage[T1]{fontenc}
\usepackage[latin9]{inputenc}
\usepackage{amsmath}
\usepackage{amssymb}
\usepackage{graphicx}

\makeatletter
\usepackage{babel}

\makeatother

\usepackage{babel}
\begin{document}

\title{Topology-Induced Symmetry Breaking for Vortex with Artificial Monopole}

\author{Zixian Zhou}
\email{zzx1313@sjtu.edu.cn}

\affiliation{Key Laboratory of Artificial Structures and Quantum Control (Ministry
of Education), Department of Physics and Astronomy, Shanghai Jiao
Tong University, Shanghai 200240, China}

\affiliation{Collaborative Innovation Center of Advanced Microstructures, Nanjing
University, Nanjing 210093, China}

\author{Zhiguo L\"{u}}
\email{zglv@sjtu.edu.cn}

\affiliation{Key Laboratory of Artificial Structures and Quantum Control (Ministry
of Education), Department of Physics and Astronomy, Shanghai Jiao
Tong University, Shanghai 200240, China}

\affiliation{Collaborative Innovation Center of Advanced Microstructures, Nanjing
University, Nanjing 210093, China}

\author{Hang Zheng}
\email{hzheng@sjtu.edu.cn}

\affiliation{Key Laboratory of Artificial Structures and Quantum Control (Ministry
of Education), Department of Physics and Astronomy, Shanghai Jiao
Tong University, Shanghai 200240, China}

\affiliation{Collaborative Innovation Center of Advanced Microstructures, Nanjing
University, Nanjing 210093, China}
\begin{abstract}
We construct an artificial $U\left(1\right)$ gauge field in the cold
atom system to form a monopole along with vortices. It is supposed
that the cold atoms are confined on a spherical surface, and two sets
of identical laser beams in the opposite propagating directions shine
on two sides of the sphere. Arbitrary Chern number $CN$, proportional
to the quantized magnetic flux, can be obtained by selecting proper
laser modes. This construction meets the condition of Chern's theorem,
so that the vortices of the atom wave function will emerge on the
sphere, whose winding number equals $CN$. It is found that a geometric
symmetry is broken spontaneously for odd $CN$, which corresponds
to a topology-induced quantum phase transition. In particular for
$CN=1$, the ground state of the cold atoms are double-degenerate
and can be applied to make a stable qubit. Since the ground-state
degeneracy is protected (led) by topology-induced symmetry breaking
against dissipation, the proposed topological structure has vast potential
in quantum storage. 
\end{abstract}

\pacs{03.75.Lm, 14.80.Hv, 42.50.Gy}

\maketitle
Gauge field plays an elementary role in various fields of physics.
The gauge intensity reflects a local geometric property, the curvature,
of intrinsic space (vector bundle) \cite{Book}. Recent researches
revealed that the gauge field can be artificially constructed in the
systems of cold atom \cite{ArtGau_org,ArtGau_imp,ArtGau_Rev} and
optical lattice \cite{Lattice1,Lattice2,Lattice3}, based on Berry's
pioneering work about the geometric phase \cite{Berry}. The idea
of the cold atom scheme resorts to a $\Lambda$-type multi-level atomic
system in laser fields \cite{ColdAtom}, such as the rubidium atoms
in Bose-Einstein condensation. In the adiabatic condition \cite{Dressed},
the atomic system generates a dressed state whose mass-center motion
is dominated by a reduced Hamiltonian with an external gauge potential
\cite{Hamilton}. It was further reported that the non-abelian gauge
\cite{NonAbel1,NonAbel2}, monopole \cite{ArtGau_imp,Monopole1,Monopole2,Monopole3},
and interacting gauge theory \cite{IntGau} can be realized by this
scheme.

Topology is closely related to the gauge field, which reflects the
global twisting of the intrinsic space. Nevertheless, there are few
studies of investigating the topological structure in the cold atom
scheme. Although the artificial magnetic field with $1/r^{2}$ behavior
has been constructed and called monopole \cite{ArtGau_imp}, it is
the $U\left(2\right)$ gauge of 3 dimension that always has a vanishing
Chern number \cite{Book} and does not generate a vortex, in other
words, a trivial topology. The Chern number as well as the winding
number of vortices are important topological invariants and extremely
useful in the topological physics such as instaton \cite{Instaton},
integer quantum Hall effect \cite{Hall}, and Kosterlitz-Thouless
transition \cite{KT}. These topological invariants are robust against
local disturbances, therefore, they have vast potential for topological
quantum storage and computation. Thus, it is interesting and significant
to see their realization in the artificial gauge system.

In this letter, we will construct an artificial gauge field with an
adjustable Chern number by the scheme of cold atom, in which the monopole
and vortex will be generated naturally. We recall Chern's theorem
\cite{Book}, also known as the Dirac quantization condition \cite{Book,DiracQuan},
that the (first) Chern number $\int_{\Sigma}B/2\pi$ ($-iB$ is the
curvature 2-form of a $U\left(1\right)$ gauge field) equals to the
winding number of vortices for the charged particle on a 2 dimensional
(2D) closed surface $\Sigma$. Then we follow this condition to propose
the cold atoms being confined on a 2D spherical surface $S^{2}$,
and couple them to a $U\left(1\right)$ gauge potential. The $U\left(1\right)$
gauge can be realized by the laser-atom coupling introduced above,
while the laser beam can only shine on a semi sphere. Thus we resort
to an identical set of laser beams with the opposite propagating direction
to cover the other part. These gauge potentials for two sides should
be equivalent along the equator, namely, differ only by a gauge transformation.
This construction is similar to the well known Dirac monopole \cite{DiracMon},
and leads to a non-zero Chern number $CN$ adjusted by the choice
of laser modes.

Chern's theorem guarantees the emergence of vortices for the cold
atoms on $S^{2}$. We solve the Schr\"{o}dinger equation for the
atoms and obtain the wave functions of the vortices. A geometric symmetry
is found spontaneously broken when $CN$ adjusted from even to odd,
which indicates the existence of a topology-induced quantum phase
transition. The ground states of the cold atoms are degenerate in
the symmetry breaking phase, in which the winding number of vortex
keeps invariant in their linear superposition. In particular, for
the case of $CN=1$, the degenerate ground states can be used to make
a stable qubit, with the location of its single vortex representing
the qubit state. This type of qubit, utilizing the ground-state degeneracy
led by topology-induced symmetry breaking, avoids the dissipation
of the traditional qubit \cite{trad_qubit}. Therefore, the topological
structure proposed in this letter has great capability in quantum
storage.

We start to construct an artificial $U\left(1\right)$ gauge field
on $S^{2}$ \cite{ArtGau_imp}. The origin Hamiltonian consists of
the kinetic energy of the mass center characterized by the Laplacian
on sphere, and a potential acting on the internal space. It takes
the form of $H_{0}=-\frac{I}{2M}\ast\textrm{d}\ast\textrm{d}+V\left(\boldsymbol{r}\right)$
in which 
\begin{equation}
V\left(\boldsymbol{r}\right)=\left(\begin{array}{ccc}
 & g_{1}\left(\boldsymbol{r}\right) & g_{2}\left(\boldsymbol{r}\right)\\
\bar{g_{1}}\left(\boldsymbol{r}\right)\\
\bar{g_{2}}\left(\boldsymbol{r}\right)
\end{array}\right),\label{eq:intrinsic}
\end{equation}
with the exterior differentiation $\textrm{d}$, Hodge star $\ast$,
$3\times3$ identity matrix $I$, atomic mass $M$, and $\hbar=1$.
This model describes a group of non-interacting cold atoms in Bose-Einstein
condensation coupled to laser light. Metric elements $g_{1,2}\left(\boldsymbol{r}\right)$
represent the laser fields that induce the Rabi oscillations between
the internal energy levels of atoms, which satisfy the Helmholtz equations.
They are usually parameterized as $g_{1}=g\cos\beta\cdot e^{i\gamma_{1}}$
and $g_{2}=g\sin\beta\cdot e^{i\gamma_{2}}$ by real functions. The
common treatment is to find the dressed state $\left|D\left(\boldsymbol{r}\right)\right\rangle $
that satisfies $V\left(\boldsymbol{r}\right)\left|D\left(\boldsymbol{r}\right)\right\rangle =0$
and to assume a heavy atomic mass $M\gg g\left(\boldsymbol{r}\right)$
to apply the Born-Oppenheimer approximation \cite{Dressed,BornOpp}.
As a result, the mass-center motion of the dressed state $\left|D\left(\boldsymbol{r}\right)\right\rangle $
is dominated by the reduced Hamiltonian $H=\left\langle D\right|H_{0}\left|D\right\rangle $
with a $U\left(1\right)$ gauge potential, given by 
\begin{equation}
H=\frac{-1}{2M}\ast\left(\textrm{d}-iA\right)\ast\left(\textrm{d}-iA\right)+\frac{W\left(\boldsymbol{r}\right)}{2M}.\label{eq:Ham}
\end{equation}
The gauge and scalar potentials are respectively given by \cite{ArtGau_imp,ArtGau_Rev}
\begin{equation}
A=-\cos^{2}\beta\textrm{d}\gamma_{1}-\sin^{2}\beta\textrm{d}\gamma_{2}+\textrm{d}\eta,\label{eq:gau}
\end{equation}
\begin{equation}
W=\cos^{2}\beta\sin^{2}\beta\left\Vert \textrm{d}\gamma_{1}-\textrm{d}\gamma_{2}\right\Vert ^{2}+\left\Vert \textrm{d}\beta\right\Vert ^{2}.\label{eq:scalar}
\end{equation}
As our scheme requires, potentials $A$ and $W$ are confined on the
sphere $r=r_{0}$, so that parameters $\beta$ and $\gamma$ are the
functions of only spherical coordinates $\theta$ and $\varphi$.
Furthermore, $A$ is undetermined with an unfixed gauge, an arbitrary
exact 1-form $\textrm{d}\eta$ on the sphere. One can design the fields
of laser to construct the required $U\left(1\right)$ gauge.

\begin{figure}
\includegraphics[width=0.9\columnwidth]{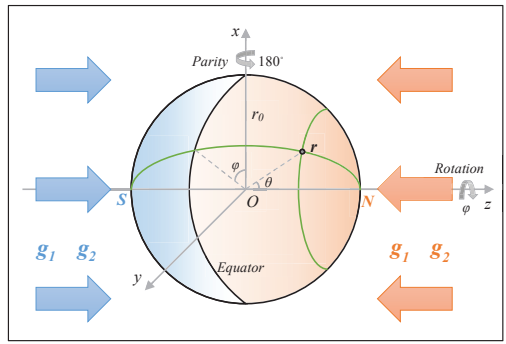}

\caption{\label{fig:Sketch}(Colors online). Configuration of the cold atom
coupled to the two sets of laser beams shining on the north and south
semi sphere, respectively. The laser beams are chosen to have parity
symmetry $P$ and $z$-axial rotation symmetry $R$.}
\end{figure}

Next we select the solutions of the Helmholtz equation to realize
the gauge with non-trivial topological structures. It has been mentioned
that we need two sets of laser beams with the opposite propagation
directions to cover the entire sphere, with the sketch presented in
Fig. \ref{fig:Sketch}. The propagating directions are chosen as $\pm z$
axis, respectively. And the two sets of laser beams are made symmetric
under the $180\textdegree$ rotation along the $x$ axis, i.e. the
2D parity for the $y-z$ coordinate $P:\left(\theta,\varphi\right)\mapsto\left(\pi-\theta,-\varphi\right)$.
For the north semi-spherical part, the laser beams $g_{1,2}\left(\boldsymbol{r}\right)$
have wave vector $k$ along the $-z$ axis, whose solution in the
spherical coordinate are given by the Gauss-Laguerre beam \cite{Laser}
\begin{equation}
g_{j}\left(r,\theta,\varphi\right)=Gc_{j}\left(\frac{\sqrt{2}r\sin\theta}{w\left(r\cos\theta\right)}\right)^{\left|l_{j}\right|}e^{-il_{j}\varphi},
\end{equation}
in which $w\left(z\right)=w_{0}\sqrt{1+z^{2}/z_{R}^{2}}$ with $z_{R}=kw_{0}^{2}/2$
the Rayleigh range and $w_{0}$ the waist size. We now assume that
the sphere locates at the origin with its radius far less than the
Rayleigh range $r_{0}\ll z_{R}$, so that we obtain $w\left(z\right)\approx w_{0}$
on the sphere. Furthermore, the common factor $G\left(r,\theta,\varphi\right)$
is a Gaussian beam but has nothing to do with the potentials, for
Eqs. (\ref{eq:gau})-(\ref{eq:scalar}) contain only the angular parameters.
Besides, $c_{j}$ are constants representing the laser intensities,
which are adjusted to $c_{1}/c_{2}=\left(\sqrt{2}r_{0}/w_{0}\right)^{\left|l_{2}\right|-\left|l_{1}\right|}$
for simplicity. Then the parameters $\beta$ and $\gamma$ are calculated
and given by $\gamma_{j}\left(\varphi\right)=-l_{j}\varphi$ and $\tan\beta\left(\theta\right)=\sin^{\left|l_{2}\right|-\left|l_{1}\right|}\theta$.
Without loosing generality, we choose $0\leq l_{1}<l_{2}$ and obtain
the potentials from Eqs. (\ref{eq:gau})-(\ref{eq:scalar}) on the
north semi sphere as follows, 
\begin{equation}
A_{N}=f\left(\theta\right)\textrm{d}\varphi+\textrm{d}\eta,
\end{equation}
\begin{equation}
W_{N}=\frac{1}{r_{0}^{2}}\frac{1+\cos^{2}\theta}{\sin2\theta}\frac{df\left(\theta\right)}{d\theta},\label{eq:W_xpr}
\end{equation}
with a continuous function defined by $f\left(\theta\right)=\left[l_{1}+l_{2}\left(\sin\theta\right)^{2l_{2}-2l_{1}}\right]/\left[1+\left(\sin\theta\right)^{2l_{2}-2l_{1}}\right]$.
Since $\textrm{d}\varphi$ has coordinate singularities at the north
and south poles, we fix the gauge as $\eta=-f\left(0\right)\varphi$
to make $A_{N}=\left[f\left(\theta\right)-f\left(0\right)\right]\textrm{d}\varphi$
analytical.

The potentials on the south semi sphere is the 2D parity transformation
$P$ of the north one, that the gauge potential reads $A_{S}=\left[f\left(\theta\right)-f\left(0\right)\right]\textrm{d}\left(-\varphi\right)$
and the scalar potential $W_{S}$ shares the same expression of Eq.
(\ref{eq:W_xpr}). Similar to the well known Dirac monopole, the gauge
potential on the equator $\left(A_{N}-A_{S}\right)_{\theta=\pi/2}=2\left[f\left(\frac{\pi}{2}\right)-f\left(0\right)\right]\textrm{d}\varphi$
should be a periodic 1-form, which requires $2\left[f\left(\frac{\pi}{2}\right)-f\left(0\right)\right]$
to be an integer \cite{DiracMon,SchrMono}. Actually this number is
the proper (first) Chern number of the $U\left(1\right)$ Berry curvature,
\begin{equation}
CN=\frac{1}{2\pi}\int_{S^{2}}\textrm{d}A=2\left[f\left(\frac{\pi}{2}\right)-f\left(0\right)\right].\label{eq:Chern}
\end{equation}
Using $f\left(\theta\right)$, we find $CN=l_{2}-l_{1}$ is indeed
an integer. We also see that arbitrary Chern number can be realized
by choosing a proper set of laser modes in this scheme. The physical
interpretation of $CN$ is given as follows: the integrand $B=\textrm{d}A$
is a differential 2-form known as the Berry curvature like a magnetic
field, then $2\pi\cdot CN$ represents the magnetic flux $\int_{S^{2}}B$.
The magnetic flux always equals to zero in the trivial case, and the
non-zero flux indicates the existence of a monopole, a non-trivial
topology. As revealed by differential geometry, a non-trivial topology
requires $A$ being defined in different areas, and jointed together
via a gauge transformation. Our designation satisfies this construction,
therefore it forms an artificial monopole.

The topology of the monopole will influence the ``charged'' particle
field. Denoting the dressed state by $\left|D\right\rangle =\int_{S^{2}}d\boldsymbol{r}\psi\left(\boldsymbol{r}\right)\left|\boldsymbol{r}\right\rangle $,
we see that the phase of the wave function $\psi\left(\boldsymbol{r}\right)$,
known as the Berry phase, is adaptive to the $U\left(1\right)$ Berry
connection $A$. By Chern's theorem, we immediately know the index
of zeros (total winding number of the Berry phase around zero points)
of $\psi$ on the sphere equals to $CN$. The zero point with a non-zero
winding number is a vortex. Therefore, in our well-constructed system,
the artificial monopole induces vortices of the cold atoms.

Then we investigate the eigenstates of the cold atoms to check the
properties of vortex, by solving the time-independent Schr\"{o}dinger
equation $H\psi=\left(\lambda/2Mr_{0}^{2}\right)\psi$. Since both
the equation and the Chern number are invariant under the 2D parity
transformation $P$, the solution space must be $P$ symmetric. Besides
this discrete symmetry, there is also the $z$-axial rotation symmetry
$R$ with generator $-i\partial_{\varphi}$, so that the spherical
variables can be separated. The basis of the solution to the Schr\"{o}dinger
equation takes the form \cite{SchrMono} 
\begin{equation}
\psi\left(\theta,\varphi\right)=\begin{cases}
\frac{1}{\sqrt{2\pi}}e^{im_{N}\varphi}\Theta\left(\theta\right), & 0\leq\theta\leq\frac{\pi}{2}\\
\frac{1}{\sqrt{2\pi}}e^{-im_{S}\varphi}\Theta\left(\theta\right), & \frac{\pi}{2}\leq\theta\leq\pi
\end{cases}.\label{eq:wave_fun}
\end{equation}
Since the gauge potential on the equator satisfies $A_{N}=A_{S}+CN\textrm{d}\varphi$,
the rule of gauge transformation requires $\psi|_{\theta=\pi^{+}/2}=e^{iCN\varphi}\psi|_{\theta=\pi^{-}/2}$,
which demands $m_{N}=CN-m_{S}$ according to Eq. (\ref{eq:wave_fun}).
Actually, this is the result of Chern's theorem as revealed later.
To obtain the equation for the latitude wave function $\Theta$, we
insert Eq. (\ref{eq:wave_fun}) to the Schr\"{o}dinger equation and
make substitution $z=\cos\theta$, arriving at 
\begin{equation}
\left[-\frac{d}{dz}\left(1-z^{2}\right)\frac{d}{dz}+\frac{F^{2}}{1-z^{2}}+r_{0}^{2}W-\lambda\right]\Theta=0,\label{eq:Theta}
\end{equation}
in which $F\left(z\right)$ is defined by $m_{S}-f\left(\pi-\theta\right)+f\left(0\right)$
at $z\in\left(-1,0\right)$ and $f\left(\theta\right)-f\left(0\right)-m_{N}$
at $z\in\left(0,1\right)$ with $F'\left(z\right)$ existing at $z=0$.
Since $F\left(z\right)$ is continuous, the wave function is second-order
differentiable. Without the artificial gauge, i.e. setting $f=0$,
Eq. (\ref{eq:Theta}) reduces to the usual Legendre equation.

\begin{figure}
\includegraphics[width=0.9\columnwidth]{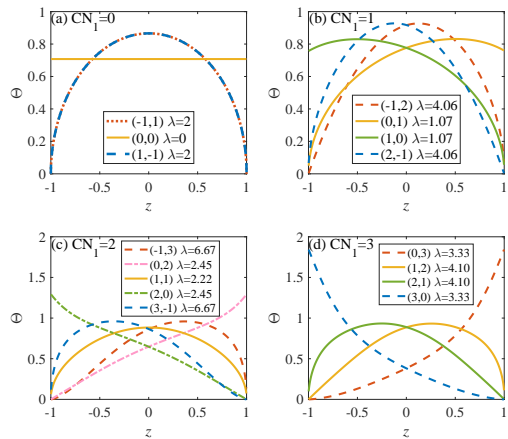}

\caption{\label{fig:wave_fun}(colors online). Eigenvalues $\lambda$, and
normalized solutions of $\Theta$ ($\int\Theta^{2}dz=1$) as functions
of $z=\cos\theta$. The Chern number $CN$ is set as 0 (a), 1 (b),
2 (c), and 3 (d), with various pairs of $\left(m_{N},m_{S}\right)$
chosen in each subfigure.}
\end{figure}

Figure \ref{fig:wave_fun} presents the behaviors of the wave functions
for various $CN$ by numerical solution, in which $l_{1}=0$ is chosen
for simplicity so that $CN=l_{2}$ from Eq. (\ref{eq:Chern}). The
most important conclusion shown in Fig. \ref{fig:wave_fun} is that
the non-zero $m_{N,S}$ induces zero of the wave function at the corresponding
pole. This happens even for the ground states, unlike the normal quantum
mechanical system whose ground state usually contains no zero. And
considering the phase around the zero reads $m_{N}\varphi$ or $m_{S}\left(-\varphi\right)$
according to Eq. (\ref{eq:wave_fun}), we find the winding number
of the phase around the north and south pole is just $m_{N}$ and
$m_{S}$ (notice the right-handed direction of $-\varphi$ corresponds
to the outside normal direction at the south pole), respectively.
Thus, that the total winding number reads $m_{N}+m_{S}=CN$ is the
result of Chern's theorem, which is consistent with the condition
of gauge transformation.

The another interesting effect presented in Fig. \ref{fig:wave_fun}
is the $RP$ symmetry $SO\left(2\right)\otimes\mathbb{Z}_{2}$ of
the Hamiltonian Eq. (\ref{eq:Ham}) is spontaneously broken for the
odd Chern number. It is seen that the ground states for the even Chern
number $CN=0,2$ (Fig. \ref{fig:wave_fun} (a) and (c)) are non-degenerate,
which certainly contains the complete symmetry; while those for the
odd Chern number $CN=1,3$ (Fig. \ref{fig:wave_fun} (b) and (d))
are double-degenerate, which means the $RP$ symmetry is broken, for
the $RP$ symmetric solution must have an even winding number. Therefore,
there is a topology-induced quantum phase transition: when the discrete
variable $CN$ varies from even to odd such as from 0 to 1, the $RP$
symmetric phase vanishes and the asymmetric phase emerges. At the
same time, the Landau order parameter, the expectation value of the
displacement $\left\langle \boldsymbol{r}\right\rangle $ for the
atoms, changes from zero to non-zero. Since the Chern number is discrete,
the energy has no derivative respect to the dependent variable, so
that this type of phase transition has no order.

\begin{figure}
\includegraphics[width=0.9\columnwidth]{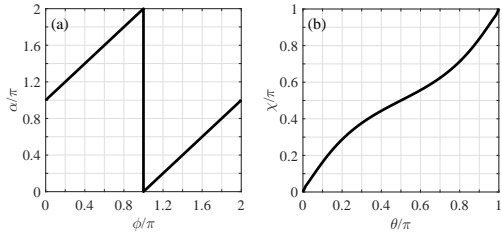}

\caption{\label{fig:qubit}$\alpha-\varphi$ relation (a) and $\chi-\theta$
relation (b) for $CN=1$. The location of the vortex is one-to-one
mapped to a point on the Bloch sphere which represents a qubit.}
\end{figure}

Here we discuss the case of $CN=1$ to reveal the $RP$ symmetry breaking
in detail. It has been shown that the ground states are double-degenerate
in this case, so that its subspace forms a Bloch sphere $\mathbb{C}P^{1}$.
The linear superposition of the ground states keeps the winding number
as 1 that is a topological invariant. Furthermore, the ground state
does not have multi vortices but only has one, which will be proved
as follows. Denoting the wave function for $\left(m_{N},m_{S}\right)=\left(1,0\right)$
by $\psi_{N}$ and that for $\left(0,1\right)$ by $\psi_{S}$, we
find their linear combination takes the form of $\psi=\psi_{N}\cos\frac{\chi}{2}+\psi_{S}e^{i\alpha}\sin\frac{\chi}{2}$
that belongs to a point on the Bloch sphere, with $\chi$ and $\alpha$
the real parameters. The location of its zero is given by the equation
$\psi=0$ which can be reduced to

\begin{equation}
\cos\frac{\chi}{2}\cdot e^{i\varphi}\Theta_{N}\left(\theta\right)+e^{i\alpha}\sin\frac{\chi}{2}\cdot\Theta_{S}\left(\theta\right)=0.
\end{equation}
From the parity symmetry $P$ we have $\Theta_{N}\left(\pi-\theta\right)=\Theta_{S}\left(\theta\right)$,
therefore, we obtain $\alpha=\varphi+\pi$ and $\tan\frac{\chi}{2}=\Theta_{N}\left(\theta\right)/\Theta_{N}\left(\pi-\theta\right)$.
The $\alpha-\varphi$ relation is quite simple as shown in Fig. \ref{fig:qubit}
(a) that results from the axial symmetry $R$. And the concrete $\chi-\theta$
relation can be obtained by solving Eq. (\ref{eq:Theta}) which is
presented in Fig. \ref{fig:qubit} (b). It is seen that both the two
relations are one-to-one, which means the zero of $\psi$ is unique,
with the winding number 1 and being a single vortex. Therefore, each
real parameter pair $\left(\chi,\alpha\right)$ on the Bloch sphere
is one-to-one mapped to the vortex location $\left(\theta,\varphi\right)$
on the sphere (diffeomorphism between $\mathbb{C}P^{1}$ and $S^{2}$).
The single vortex may appear at the poles, keeping the $R$ symmetry
but breaking the $P$ symmetry; or it locates at the equator to maintain
the $P$ symmetry but sacrifices the $R$ symmetry. All in all, the
$RP$ symmetry must be broken for $CN=1$, which still keeps correct
for the case of higher odd Chern number.

What is the application of this topological structure? The degenerate
ground states for $CN=1$ are suitable for making a qubit. The superposition
state of $\psi_{N}$ and $\psi_{S}$, locating on the Bloch sphere,
represents a bit of quantum information. The qubit state can be read
by measuring the location of the single vortex formed by a group of
cold atoms in condensation. We know that the intensity of linear response
is proportional to the density of matter, thus the vortex as the zero
point can be detected by the response signal to a weak external field
\cite{MassCenter}. Next, we point out that this qubit is protected
against dissipation. From the beginning if the dissipation rate of
the internal levels for potential Eq. (\ref{eq:intrinsic}) is much
less than the light intensity $g$, the dissipation will be negligible
and the dressed state will have an appreciable lifetime. Then suppose
the $RP$ symmetry and topological structures proposed in this letter
are realized, then the degenerate ground states, resulted from topology-induced
symmetry breaking, evidently have no dissipation. Thus, the qubit
is stable against dissipation. Besides, the required topology is determined
by the lasers which can be fined tuned and keep stable. And suppose
there is small deficiency of the required symmetry, one can still
adjust the light field to compensate the asymmetry and restrict the
degeneracy splitting. In short, our topological scheme provides an
ideal realization of quantum storage device.

In summary, we construct an artificial $U\left(1\right)$ gauge potential
in the cold atom system on a spherical surface to form a monopole
and vortex. The monopole with arbitrary Chern number can be obtained
by selecting proper coupling laser beams, and it induces the vortex
of the atoms appearing on the sphere via Chern's theorem. Then the
Schr\"{o}dinger equation is solved to check the total winding number
of vortices equal $CN$. For the odd $CN$, it is found that the geometric
$RP$ symmetry is spontaneously broken, for the ground states of atoms
are double-degenerate. This topology-induced phase transition has
a discrete dependent variable $CN$, without a transition order. In
particular for $CN=1$, it is shown that the linear coefficient of
superposition ground state is one-to-one mapped to the location of
single vortex. These degenerate ground states are suitable for making
a qubit whose state can be read by detecting the vortex location.
Besides, the ground-state degeneracy led by topology-induced symmetry
breaking is stable against dissipation. It is interesting to see the
realization of the proposed topological structure in the cold atom
experiments, which possesses an application significance in quantum
storage.

This work was supported by the National Natural Science Foundation
of China under Grants No. 11374208 and No. 11474200.

\end{document}